\documentclass[12pt]{article}
\usepackage{epsfig}
\usepackage{palatino}
\usepackage[latin1]{inputenc}
\usepackage{epsf}
\usepackage{graphicx,psfrag,color,pstcol,pst-grad}
\usepackage{amsmath,amssymb}
\usepackage{latexsym}
\usepackage{calc}
\def\be{\begin{equation}} \def\ee{\end{equation}} \def\bea{\begin{eqnarray}} 
\def\eea{\end{eqnarray}} \def\nnb{\nonumber} 
\def\gat{\tilde{g}_A}

\begin{document}

\begin{center}
{\large\bf Monte Carlo Study of the abBA Experiment: \\
Detector Response and Physics Analysis}\\[0.5cm]
{E.~Frle\v z},  for the abBA Collaboration\\[0.3cm]

Department of Physics, University of Virginia,\\
Charlottesville, VA 22904-4714, USA\\[0.3cm]
\end{center}

\begin{abstract}
The abBA collaboration proposes to conduct a comprehensive program 
of precise measurements of neutron $\beta$-decay coefficients 
$a$ (the correlation between the neutrino momentum and the decay electron
momentum), $b$ (the electron energy spectral distortion term), $A$
(the correlation between the neutron spin and the decay electron momentum), 
and $B$ (the correlation between the neutron spin and the decay neutrino 
momentum) at a cold neutron beam facility. We have used a GEANT4-based 
code to simulate the propagation of decay electrons and protons in the
electromagnetic spectrometer and study the energy and timing response 
of a pair of Silicon detectors. We used these results to examine
systematic effects and find the uncertainties with which 
the physics parameters $a$, $b$, $A$, and $B$ can be extracted
from an over-determined experimental data set.


\end{abstract}

\section{Introduction}\label{sec:int}
\medskip

The abBA collaboration is proposing to perform a measurement of 
a ``complete set'' of correlations in the neutron $\beta$-decay
using the same apparatus, and improve the precision of 
the correlation coefficients $a$, $b$, $A$, and $B$ by up to 
an order of magnitude. 

GEANT4 is a general-purpose software package for simulation of 
the passage of particles through matter that provides a complete 
set of tools for all domains of detector simulation~\cite{G4}.
In particular, the GEANT4 toolkit currently provides particle tracking
in non-uniform magnetic and electric fields and
handles combined electromagnetic fields transparently~\cite{Wri02}.
The GEANT4 Low Energy Electromagnetic Physics group validates
the low energy electromagnetic processes for electrons down to 
250\,eV~\cite{low}. 

In this report we describe a GEANT4 simulation of the abBA
spectrometer and outline the algorithm for the extraction of the physics
decay parameters.

\section{abBA Detector Geometry}\label{sec:geo}
\medskip
 
\begin{figure} [!tpb]      
\begin{center}
\includegraphics[width=12cm]{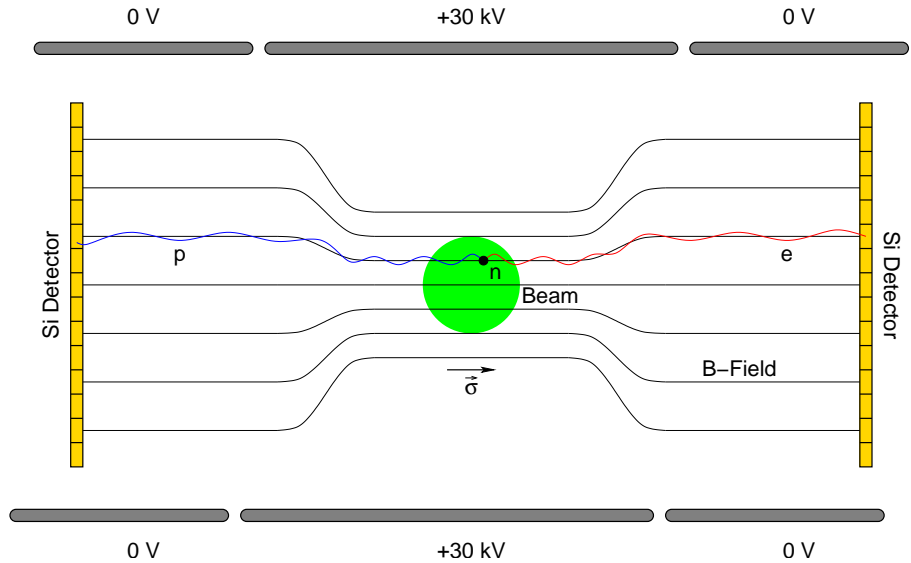} 
\end{center}
\caption{The schematic drawing of the main elements of the abBA spectrometer.}
\label{det}
\end{figure}

In the tentative design of the abBA spectrometer the decay particles
(electrons and protons) are guided by the electric and magnetic fields
and interact only in sensitive detectors thus avoiding the energy losses
and scatterings in apertures, grids, or windows~\cite{abba}. 

The simplified geometry of the detector defines two sensitive planar Silicon 
detectors with a $100\times100$\,mm$^2$ area and a 2\,mm thickness. The two 
Si detectors are separated by 4 meters. The coordinate system is defined 
with the Si detectors at $x_{1,2}=\pm 4$ meters (Fig.~\ref{det}).

A passive solenoid magnet is placed around the decay region, with its axis of
symmetry perpendicular to the incident neutron beam, along the $x$ coordinate. 
The 3\,m long magnet with a 0.8\,m radius can produce a 4\,T central
magnetic field that decreases to 1\,T at the detector positions,
thus guiding charged particles from the decay region 
to the Si detectors. A tubular electrode held at $\sim$30\,kV accelerates
the protons so they can be detected in a Silicon detector.
\bigskip                

\section{Magnetic Field}\label{sec:field}
\medskip
 
The magnetic field along the $x$ axis of the detector solenoid is given by:
\begin{displaymath}
B(x,\rho=0)=\frac{ 2\pi NI}{c }\left(
{ {L-x}\over {\sqrt{R^2+(L-x)^2}} } +
{ {x}\over {\sqrt{R^2+x^2}} } \right),
\end{displaymath}
where $L$ is the length of the solenoid, $R$ its radius,
and $x$ the axial coordinate. Meanwhile, $N$ denotes the number of turns
per unit length, and $I$ is the electrical current in the closely 
wound cylindrical coil.
 
Thanks to axial symmetry, the magnetic field off-axis,
outside its sources, can be represented in terms of the
magnetic field strength $B(x,0)$ along the axis:
\begin{eqnarray}
B_x(x,\rho) && =-{ {\partial\phi}\over {\partial x} }
 = \sum_{n=0}^\infty { {(-1)^n}\over {(n!)^2} }
B^{(2n)}(x) \left( {\rho\over 2}  \right)^{2n} \nonumber
 = B(x)- { {\rho^2}\over 2 }B^{\prime\prime}(x,0)+ \cdots,
\end{eqnarray}
and
\begin{eqnarray}
B_\rho(x,\rho) && =-{ {\partial\phi}\over {\partial \rho} }
 = \sum_{n=1}^\infty { {(-1)^n}\over {(n-1)!n!} }
B^{(2n-1)}(x) \left( {\rho\over 2}  \right)^{2n-1} \nonumber
  = - { {\rho}\over 2 }B^{\prime}(x,0)+ \cdots,
\end{eqnarray}
where $\phi$ is the magnetic scalar potential and $\rho=\sqrt{y^2+z^2}$ 
is the axial radius coordinate. These fields have been programmed into 
the GEANT4 user routine.
\bigskip
    
\section{Event Generator}\label{sec:event}
\medskip 
 
The electrons from the neutron $\beta$-decay are generated
from $5\times 5\times 5\,$mm$^3$ central volume with the relativistic
differential decay rate given by~\cite{jac57}:           
\begin{displaymath}
\frac{d\Gamma}{dE_e d\Omega_{p_e}d\Omega_{p_\nu}} = 
\frac{(G_FV_{ud})^2}{(2\pi)^5}
\frac{ F(E_e) \; |\vec{p}_e| \; E_\nu}{ m_n \; 
[E_p+E_\nu+E_e\; (\vec{\beta}\cdot\hat{p}_\nu)]} \vert M\vert ^2,
\end{displaymath}
where $E_e$ and $\vec{p}_e$ ($E_\nu$ and $\vec{p}_\nu$) are the 
electron (neutrino) energy and momentum, $m_n$ is the neutron mass,
$G_F$ is the Fermi constant, $V_{ud}$ is the Cabbibo-Kobayashi-Maskawa
matrix element,
$\vec{\beta}=\vec{p}_e/E_e$, and $F(E_e)$ is the Fermi function
that describes the interaction of the electron and the recoil proton.

\begin{figure}[!tpb]
\vglue -2.5cm
\begin{center} 
\includegraphics[width=14cm]{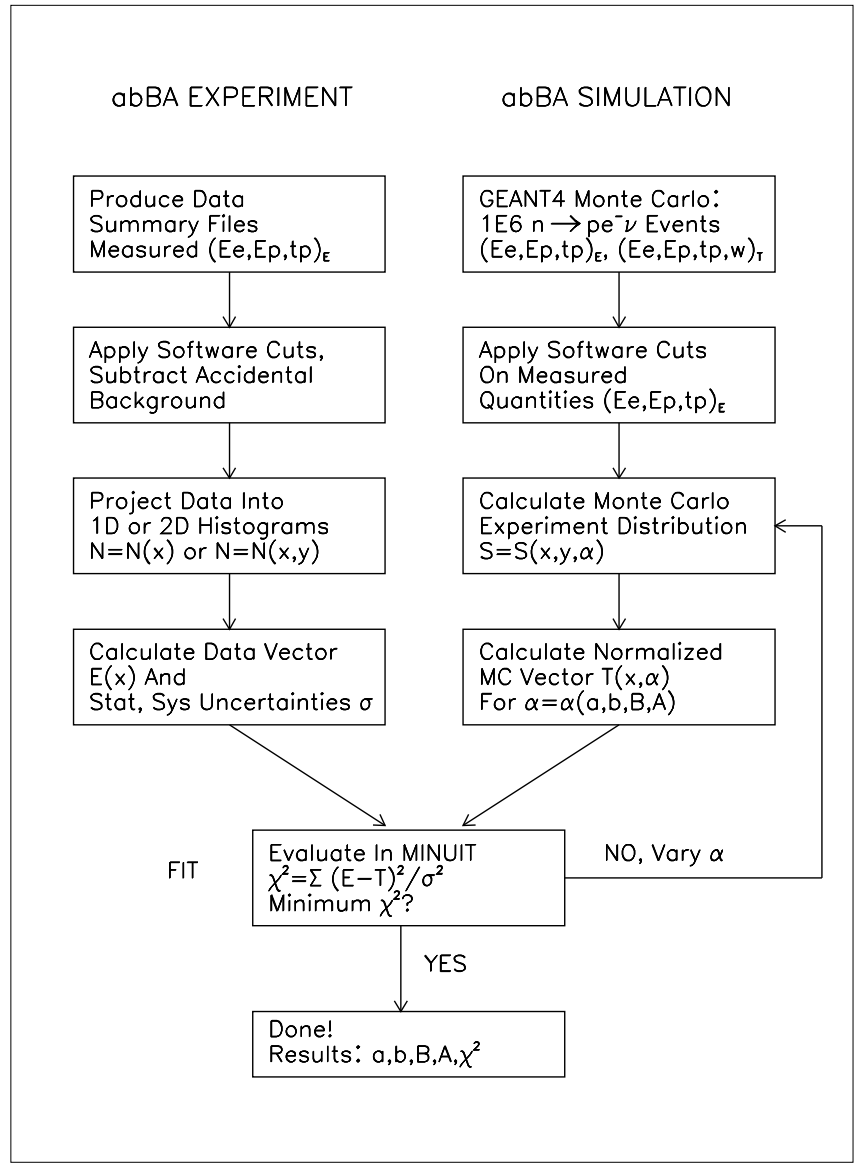}
\end{center}
\vglue -1.5cm
\caption{A flow diagram of the analysis. Monte Carlo histograms
are recalculated in each step of {\tt MINUIT} minimization. 
The ``experiment'' depends only on measured quantities (subscript $E$),
while the MC column depends both on generated variables (subscript
$T$) and simulated detector response (subscript $E$).}
\label{alg}
\end{figure}

The transition matrix element squared $ \vert M\vert ^2$ is given by
\bea 
\lefteqn{ |M|^2 
= m_nm_pE_eE_\nu 
\left( 1 +\frac{\alpha}{2\pi} \; e_V^R \right)
\left( 1 +\frac{\alpha}{2\pi} \; \delta_\alpha^{(1)} \right)
 }
\nnb \\ && \times C_0(E_e) (1+3 \gat^2) 
\left\{
1+ \left( 1 +\frac{\alpha}{2\pi} \; \delta_\alpha^{(2)} \right) 
C_1(E_e) \vec{\beta}\cdot\hat{p}_\nu+b\left( \frac{m_e}{E_e}\right)
\right. \nnb \\ && \left.
+ \left( 1 +\frac{\alpha}{2\pi} \; \delta_\alpha^{(2)} \right)
[C_2(E_e)+C_3(E_e) \vec{\beta}\cdot\hat{p}_\nu ]
\hat{n}\cdot\vec{\beta} \right. \nnb \\ && \left.
+[C_4(E_e)+C_5(E_e) \vec{\beta}\cdot\hat{p}_\nu ]
\hat{n}\cdot\hat{p}_\nu
\right\} \nnb,
\eea 
where $m_p$ is the proton mass, $\alpha$ is the fine structure constant,
$e_V^R$ is a low energy constant, $\delta_\alpha$'s are model-independent 
radiative corrections, $\gat$ is the axial coupling constant,
and the correlation coefficients $a$, $A$, $B$ are incorporated
into the recoil corrections $C_i(E_e)$~\cite{and04}.

\section{Results and Conclusions}\label{sec:res}
\medskip

A GEANT4 simulation of abBA detector energy and timing response was performed
for $10^6$ neutron $\beta$-decays. We used the values of the correlation 
coefficients from Ref.~\cite{Glu95} ($a=-0.1039$, $b=0$, $A=-0.1161$, 
$B=0.9878$). For each event we recorded the neutron polarization,
generated momenta of the final state particles and measured energy depositions
and timing hits in the Silicon detectors.

The separate GEANT4 run which included systematic effects (energy and timing 
resolutions of Si detectors, detector calibration uncertainties,
detector response nonlinearities, magnetic field inhomogeneities, 
neutron polarization uncertainty, etc.) was used to simulate the experimental data.
(``MC data''). The flow chart of the physics analysis is summarized in Fig.~\ref{alg}.
The correlation coefficients and their fitted uncertainties are extracted using 
the standard {\tt MINUIT} code~\cite{jam89}.

\begin{figure}[!tpb]
\vglue -2cm
\begin{center}              
\includegraphics[width=14cm]{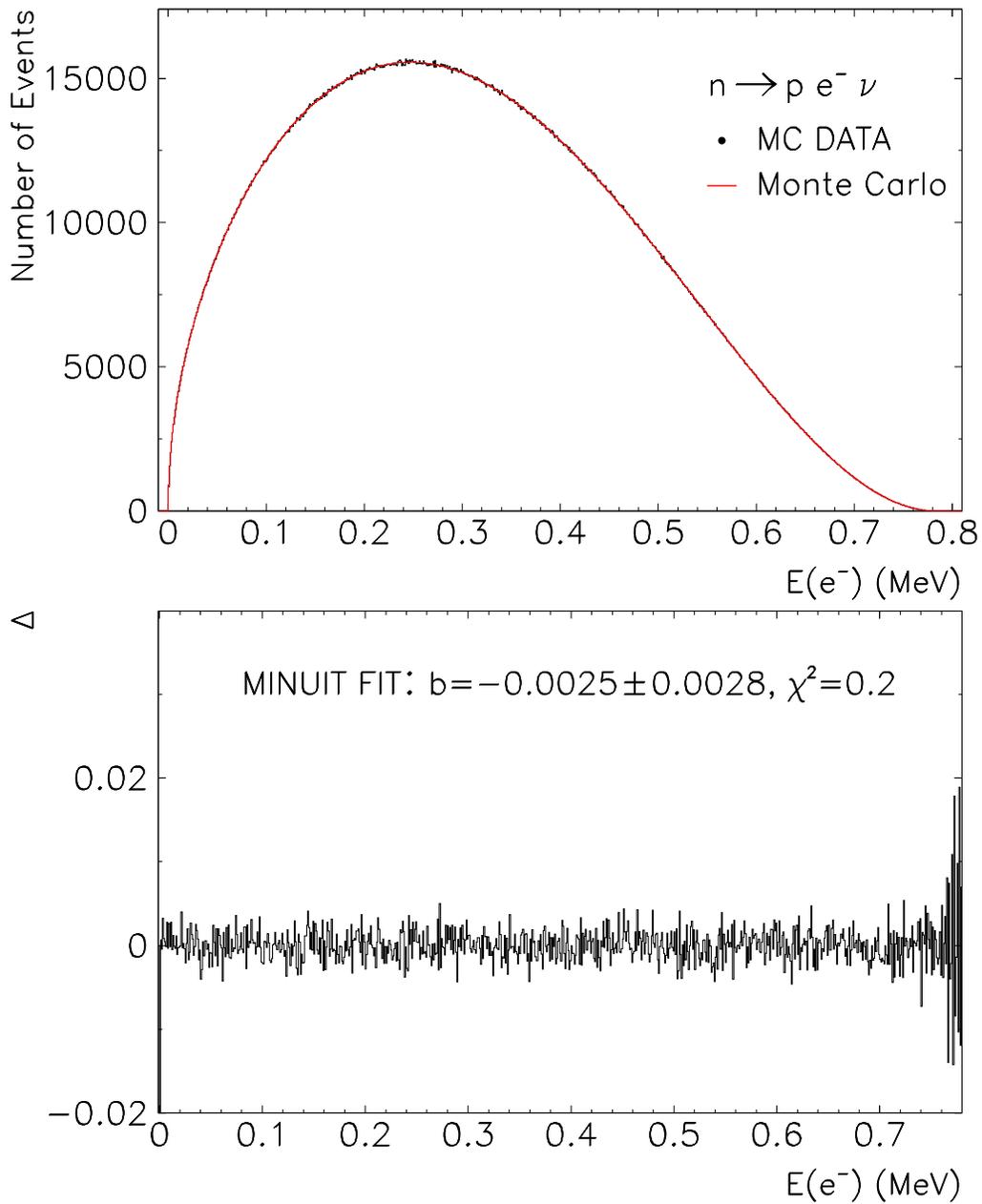} 
\end{center}
\caption{Monte Carlo energy spectrum of neutron $\beta$-decay electron (top)
and fractional differences $\Delta={(exp-the)/(exp+the)}$ between 
the simulated and ``MC data'' spectrum (bottom).}
\label{sa}
\end{figure}

\begin{figure}[!tpb]
\vglue -2cm
\begin{center}              
\includegraphics[width=14cm]{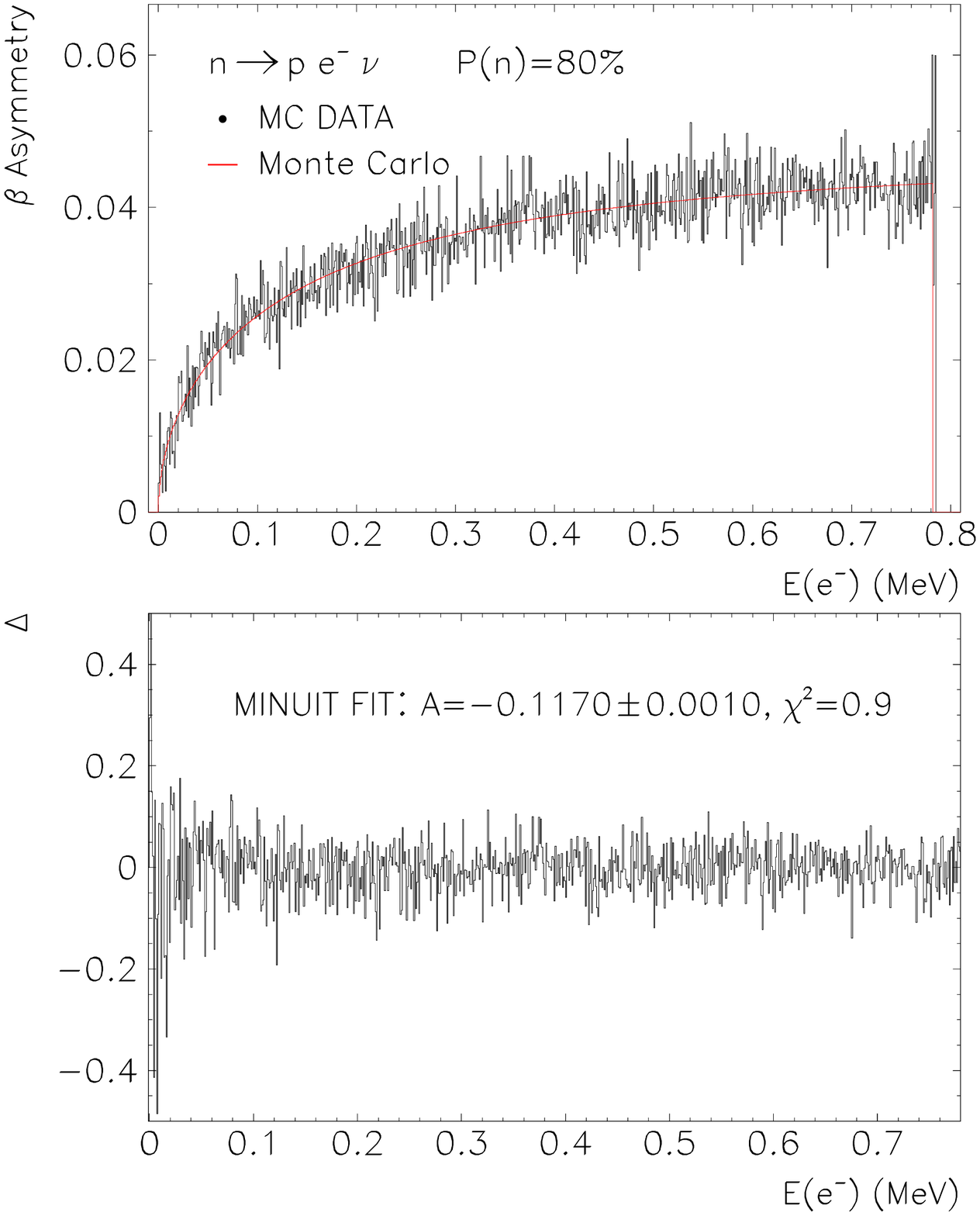} 
\end{center}
\caption{Monte Carlo histogram of the $\beta$ asymmetry (top)
and fractional differences between the simulated and ``MC data'' spectrum 
as a function of the electron energy (bottom).}
\label{asy}
\end{figure}

We present two example results: (i) the extraction of the parameter $a$ with unpolarized 
neutron beam in Fig.~\ref{sa}, and (ii) the asymmetry coefficient $A$ for 80\,\% 
polarized neutron beam in Fig.~\ref{asy}. At the current stage of development,
a GEANT4 simulation limited to $10^6$ neutron decay events and $10^6$ MC data events, 
runs 24 CPU hours on a 1\,GHz Linux computer. Given the limited event statistics, 
analysis of MC data results in the coefficient $b=-0. 0025\pm 0.0028$ and the coefficient
$A=-0.1170\pm 0.0010$, where statistical and systematic uncertainties are combined. 
The code will be made faster by using the adiabatic invariants for charged particle 
tracking in the electromagnetic field, which will markedly improve the uncertainties
of our method.

\clearpage
\vspace*{\stretch{1}}

\vfill
\vspace*{\stretch{2}}

\end{document}